\documentclass[epj,nopacs]{svjour} 
 
\usepackage{latexsym} 
\usepackage{epsfig}
\usepackage{amsmath}
\usepackage{cite}
 
\begin{document} 
 
\title{Chiral quark model analysis \\ of nucleon quark sea isospin
asymmetry and spin polarization} 

\author{T. Ohlsson\thanks{\email{tommy@theophys.kth.se}}, H.
Snellman\thanks{\email{snell@theophys.kth.se}}}

\institute{Theoretical Physics, Department of
Physics, Royal Institute of Technology, SE-100 44 Stockholm, Sweden} 
 
\date{Received: 17 June 1998}

\abstract{We analyze recent measurements of the nucleon quark sea isospin
asymmetry in terms of the chiral quark model.  The new measurements
indicate that the SU(3) model with modest symmetry breaking and no
$\eta'$ Goldstone boson gives a satisfactory description of data.  We
also discuss the matching parameter for the axial-vector current.
Finally, we analyze the nucleon quark spin polarization measurements
directly in the chiral quark model without using any SU(3) symmetry
assumption on the hyperon axial-vector form factors.  The new data
indicate that the chiral quark model gives a remarkably good and
consistent description of all low energy baryon measurements.}
 
\authorrunning{T. Ohlsson \and H. Snellman}
\titlerunning{Chiral quark model analysis}
\maketitle

\section{Introduction}
\label{sec:intro}

The parameterization of low energy hadron structure by the chiral
quark model ($\chi$QM), suggested by Manohar and Georgi \cite{mano84}
have enjoyed rising interest recently
\cite{eich92,chen95,song97,webe97,webe972,chen98,lind98,song98,lind982,chen982,webe98,ohls98,song982,song983}.
Especially, the emission and absorption of pseudoscalar Goldstone
bosons (GBs) from the quarks lead to a spin depolarization that seems
consistent with measurements of the quark spin polarization in
nucleons, axial-vector form factors and magnetic moments.

The $\chi$QM Lagrangian for the quark-GB interaction can be written, to
lowest order, as
\begin{equation}
{\cal L} = g_8 \bar{\bf q} \Phi \gamma^{5} {\bf q},
\label{eq:lag}
\end{equation}
where $g_8$ is a coupling constant,
$$
{\bf q} = \left( \begin{array}{c} u \\ d \\ s \end{array} \right),
\; \mbox{and} \; \Phi = \left( \begin{array}{ccc}
\frac{\pi^0}{\sqrt{2}}+ \beta \frac{\eta}{\sqrt{6}} & \pi^+ &
\alpha K^+ \\ \pi^- & -\frac{\pi^0}{\sqrt{2}}+ \beta
\frac{\eta}{\sqrt{6}} & \alpha K^0 \\ \alpha K^- &
\alpha\bar{K}^0 & - \beta \frac{2\eta}{\sqrt{6}} \end{array} \right)
$$
The GBs of the $\chi$QM are here denoted by the $0^{-}$ meson names
$\pi,K,\eta,\eta'$, as is usually done.  We have introduced two SU(3)
symmetry breaking parameters, $\alpha$ and $\beta$, which allow for
different strengths of production of GBs containing strange quarks.

To account for the quark sea isospin asymmetry as measured by the NA51
Collaboration \cite{bald94} and the New Muon Collaboration (NMC)
\cite{amau91,arne94}, Cheng and Li \cite{chen95} suggested the
introduction of a broken U(3) symmetric model with nine GBs.  The
ninth GB, the $\eta'$, should couple with a relative strength $\zeta$,
that is different from the strength of the octet GBs (when $\alpha =
1$ and $\beta = 1$). The interaction Lagrangian has the form ${\cal
L'} = g_8 \zeta \frac{1}{\sqrt{3}} \bar{\bf q} \eta' \gamma^{5} {\bf q}$.

The probability of transforming a quark with spin up by one
interaction can then be expressed by the functions
\begin{eqnarray}
\vert\psi(u^\uparrow)\vert^2 &=& \tfrac{1}{6} a (3 + \beta^2 + 2
\zeta^2) \hat{u}^\downarrow + a \hat{d}^\downarrow + a
\alpha^{2} \hat{s}^\downarrow, \\
\vert\psi(d^\uparrow)\vert^2 &=& a \hat{u}^\downarrow +
\tfrac{1}{6} a (3 + \beta^2 + 2 \zeta^2) \hat{d}^\downarrow + a
\alpha^{2} \hat{s}^\downarrow, \\ \vert\psi(s^\uparrow)\vert^2
&=& a \alpha^{2} \hat{u}^\downarrow + a \alpha^{2}
\hat{d}^\downarrow + \tfrac{1}{3} a (2 \beta^2 +\zeta^2)
\hat{s}^\downarrow.
\end{eqnarray}
The coefficient of a quark $\hat{q}^\downarrow$ is the transition
probability to $q^{\downarrow}$. The parameter $a$ ($a \propto g_8^2$)
measures the probability of emission of a GB from a quark and is
sometimes called the fluctuation (or probability) parameter.

The NA51 experiment gave for the quark sea isospin asymmetry the value
$\bar{u}/\bar{d} = 0.51 \pm 0.09$ at $x = 0.18$ \cite{bald94}.  This
excludes that the value $\zeta=0$ can be used, at least for $\beta \simeq 1$.

However, recently there has been a remeasurement of both the
$\bar{u}/\bar{d}$ asymmetry as well as of the $\bar{u} - \bar{d}$
asymmetry by the NuSea Collaboration\footnote{Fermilab E866/NuSea
Collaboration} \cite{hawk98,peng98}.  Their data differ substantially
from the earlier measurements.  It therefore seems appropriate to
re-evaluate the range of the parameters of the $\chi$QM in view of the
new data.  In this paper, we undertake such a re-evaluation in the
framework of the broken SU(3) symmetric $\chi$QM, and confront the
model with the spin polarization measurements of the nucleons.

The new measurements lead to the conclusion that there is no need for
a ninth GB.  Thus we can set $\zeta = 0$.  The magnetic moment data
show that there is no need to distinguish the $\eta$ GB from the $\pi$
GBs.  Thus only two parameters, $a$ and $\alpha$, remain in the model.
We find the best fit for those parameters, and show that these values
are in agreement with the matching parameter $g_{a}=1$, in accordance
with the analysis of Weinberg \cite{wein90,wein91} and with the
nucleon axial-vector from factor $g_A^{np}$. Since the model
parameters can be determined from nucleon data alone, we can study the
nucleon quark spin polarization measurements in this model without
introducing any assumption on the axial-vector form factor
$a_{8}$. This allows in principle the possibility to study the gluon
polarization in the nucleon.

The outline of our paper is as follows.  In Sect.~\ref{sec:sea} we
discuss the quark sea isospin asymmetry and also the so called
matching parameter $g_{a}$ introduced by Manohar and Georgi
\cite{mano84}.  This parameter is closely related to the integral
expression for the parameter $a$ \cite{eich92}.  In
Sect.~\ref{sec:nucleon} we investigate the parameter space for the
$\chi$QM using the neutron-proton axial-vector from factor $g_A^{np}$
and confront the model with the best measurements for the quark spin
structure functions. This eventually leads to a determination of the
nucleon gluon spin polarization.  Finally, in
Sect.~\ref{sec:summary}, we present a summary of our analysis and
also the main conclusions.

\section{Nucleon quark sea isospin asymmetry}
\label{sec:sea}

\begin{figure}
\epsfig{figure=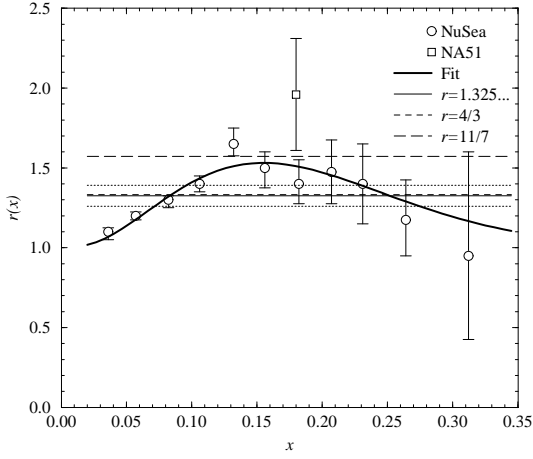,height=7cm}
\caption{The $\bar{d}/\bar{u}$ ratio at $Q = 7.35 \, {\rm GeV}$. $r(x)
\equiv \bar{d}(x)/\bar{u}(x)$, where $0.02 < x < 0.345$. The 11 data
points marked with $\circ$ were obtained by the NuSea Collaboration
\protect\cite{hawk98,peng98} and the data point marked with $\Box$ was
obtained by the NA51 Collaboration \protect\cite{bald94}. The thick
solid curve was obtained by the NuSea Collaboration by fitting to the
11 experimental data points (not including the data point from the
NA51 experiment). The analytical expression for the fitted curve is $1
+ 1120 x^{2.75} (1-x)^{15}$. The thin solid line is the average
$\bar{d}/\bar{u}$ ratio, which is about 1.325, and the dotted lines
are the corresponding 5 \% errors. The dashed line shows the
$\bar{d}/\bar{u}$ ratio for the SU(3) $\chi$QM and the long dashed
line shows the $\bar{d}/\bar{u}$ ratio for the SU(2) $\chi$QM ({\it i.e.}
the $\chi$QM with just pions as GBs)}
\label{fig:dbardubar}
\end{figure}

As mentioned above, the quark sea isospin asymmetry in the nucleon has
previously been measured by the NA51 experiment to be $\bar{u}/\bar{d}
= 0.51 \pm 0.09$ at $x=0.18$ \cite{bald94}.  The new measurement by
the NuSea Collaboration \cite{hawk98,peng98} gives this asymmetry for
a range of $x$-values.  Their value for the $\bar{u}/\bar{d}$
asymmetry at $Q=7.35 \, {\rm GeV}$ can be accurately fitted by the
formula
\begin{equation}
\bar{d}(x)/\bar{u}(x) = 1 + 1120 x^{2.75} (1-x)^{15}
\label{eq:d/u(x)}
\end{equation}
in the region $0.02 < x < 0.345$ (see Fig.~\ref{fig:dbardubar}).
Lacking independent measurements of $\bar{u}(x)$ and $\bar{d}(x)$, we
will here define the $\bar{u}/\bar{d}$ asymmetry as
\begin{equation}
\bar{u}/\bar{d} \equiv \left( \int_0^1 \bar{d}(x)/\bar{u}(x) \, dx
\right)^{-1}.
\end{equation}
By integrating (\ref{eq:d/u(x)}) over the given region, we obtain
the asymmetry as
$$
\bar{d}/\bar{u} \simeq \frac{\int_{0.02}^{0.345} \bar{d}(x)/\bar{u}(x) \,
dx}{0.345 - 0.02} \approx 1.325,
$$
whence $\bar{u}/\bar{d} \approx 0.755$\footnote{Another possible
definition of the $\bar{u}/\bar{d}$ asymmetry is $\bar{u}/\bar{d}
\equiv \int_0^1 \left(\bar{d}(x)/\bar{u}(x)\right)^{-1} \, dx$. This
definition leads to $\bar{u}/\bar{d} \approx 0.765$, which is almost
the same as $\bar{u}/\bar{d} \approx 0.755$.}.

In the $\chi$QM, this asymmetry is given by \cite{lind98}
\begin{equation}
\bar{u}/\bar{d} = \frac{21 + 2\xi + \xi^2}{33 - 2\xi + \xi^2},
\label{eq:u/d}
\end{equation}
where $\xi \equiv 2 \zeta + \beta$.  Comparing this with the new
result above shows that the asymmetry is compatible with the parameter
$\xi = 1$ for which
$$
\bar{u}/\bar{d} = \frac{3}{4} = 0.75.
$$
The value $\xi = 2 \zeta + \beta = 1$ implies that for a value of
$\beta$ around 1, we can put $\zeta=0$, corresponding to complete
$\eta'$ suppression. Evidence for a strong $\eta'$ suppression has
also been suggested by Song \cite{song98} and is discussed in
\cite{webe97,chen98,lind982}. The value $\zeta = 0$ is welcome,
since our understanding from QCD concerning the role of the $\eta'$ is
that it is not a GB, but related to instantons via the axial anomaly
\cite{thoo76}.

In Fig.~\ref{fig:dbardubar} we see that the $\bar d / \bar u$
asymmetry for the SU(3) $\chi$QM coincides remarkably well with the
experimental result.  In the paper by the NuSea Collaboration
\cite{peng98}, it is indicated in their Fig.~3, and commented on in
the text, that the $\chi$QM is not compatible with data.  However,
their quoted value of $11/7$ for the $\bar d/ \bar u$ asymmetry in the
$\chi$QM is relevant only for the SU(2) $\chi$QM (corresponding to
$\beta = 0$ and $\zeta = 0$, {\it i.e.} $\xi = 0$ in (\ref{eq:u/d})),
and not for the SU(3) $\chi$QM.  We should thus instead interpret the
new result as an experimental verification of the presence of $\eta$
GBs in the $\chi$QM.

We next study the quark sea isospin asymmetry $\bar{u} - \bar{d}$
appearing in the Gottfried sum-rule \cite{gott67}.  Here the previous
value is $\bar{u} - \bar{d} = - 0.15 \pm 0.04$, which was obtained by
the NMC \cite{amau91,arne94}.  In \cite{peng98}, the NuSea
Collaboration has measured this to be $\bar{u} - \bar{d} = -0.100 \pm
0.024$ at $Q = 7.35 \, {\rm GeV}$. This is $2/3$ of the value deduced
by the NMC.  We can now use this new measurement to estimate the
parameter $a$ in the $\chi$QM.  We will use the parameter values
$\beta = 1$ and $\zeta = 0$, {\it i.e.} $\xi = 1$.  From the formula of the
$\bar{u} - \bar{d}$ asymmetry in the $\chi$QM \cite{lind98}
\begin{equation}
\bar{u} - \bar{d} = a \left(\frac{2 \zeta + \beta}{3} - 1\right) = a
\left(\frac{\xi}{3} - 1\right),
\label{eq:u-d}
\end{equation}
we then obtain $a = 0.150 \pm 0.036$.

In the $\chi$QM, $\bar{u}/\bar{d}$ and $\bar{u} - \bar{d}$ are both
independent of $\alpha$. The parameters $\beta$ and $\zeta$ determine
the relative mixing of the isospin triplet $\bar{\pi}$ and the isospin
singlets $\eta$ and $\eta'$.

The probability parameter $a$ has also been calculated using the chiral field
theory approach \cite{eich92}. The result is
\begin{eqnarray}
a &=& \frac{g_8^2}{32 \pi^2} \int_0^1 \theta(\Lambda_{\chi{\rm SB}}^2 -
\tau(z)) z \nonumber \\
&\times& \left\{ \ln \frac{\Lambda_{\chi {\rm SB}}^2 +
m_\pi^2}{\tau(z) + m_\pi^2}
+ m_\pi^2 \left[ \frac{1}{\Lambda_{\chi {\rm SB}}^2 + m_\pi^2} -
\frac{1}{\tau(z) + m_\pi^2} \right] \right\} \nonumber \\
&\times& dz,
\label{eq:a}
\end{eqnarray}
where $g_8 \equiv 2m g_a/f_\pi$, $\tau(z) \equiv m^2 z^2/(1-z)$, and
$\theta$ is the Heaviside function. The parameters $m$, $m_\pi$,
$f_\pi$, and $\Lambda_{\chi {\rm SB}}$ are the quark mass, the GB
mass, the pseudoscalar decay constant, and the chiral symmetry
breaking scale, respectively.  In the integral expression above for
the parameter $a$, the so called ``matching parameter'', $g_a$,
introduced by Manohar and Georgi \cite{mano84}, occurs. The parameter
$g_a$ is sometimes also called the quark axial-vector current coupling
constant.  Using the value $a=0.15$ and the parameters $m = 363 \,
{\rm MeV}$, $m_\pi = 140 \, {\rm MeV}$, $f_\pi = 93 \, {\rm MeV}$,
$\Lambda_{\chi {\rm SB}} \simeq 4 \pi f_\pi = 1169 \, {\rm MeV}$, one
obtains $g_a \approx 0.987 \sim 1$ from solving (\ref{eq:a}). This
is in good agreement with the arguments of Weinberg
\cite{wein90,wein91}, that to lowest order in $1/N_c$, where $N_c$ is
the number of colors, the value of $g_a$ should be 1.

\begin{figure}
\epsfig{figure=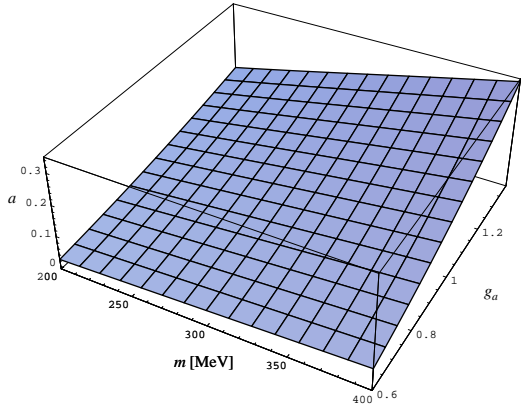,height=7cm}
\caption{The probability parameter $a$ plotted against the quark mass
$m$ and the matching parameter (the quark axial-vector current coupling
constant) $g_a$}
\label{fig:a_m_ga}
\end{figure}

\begin{figure}
\epsfig{figure=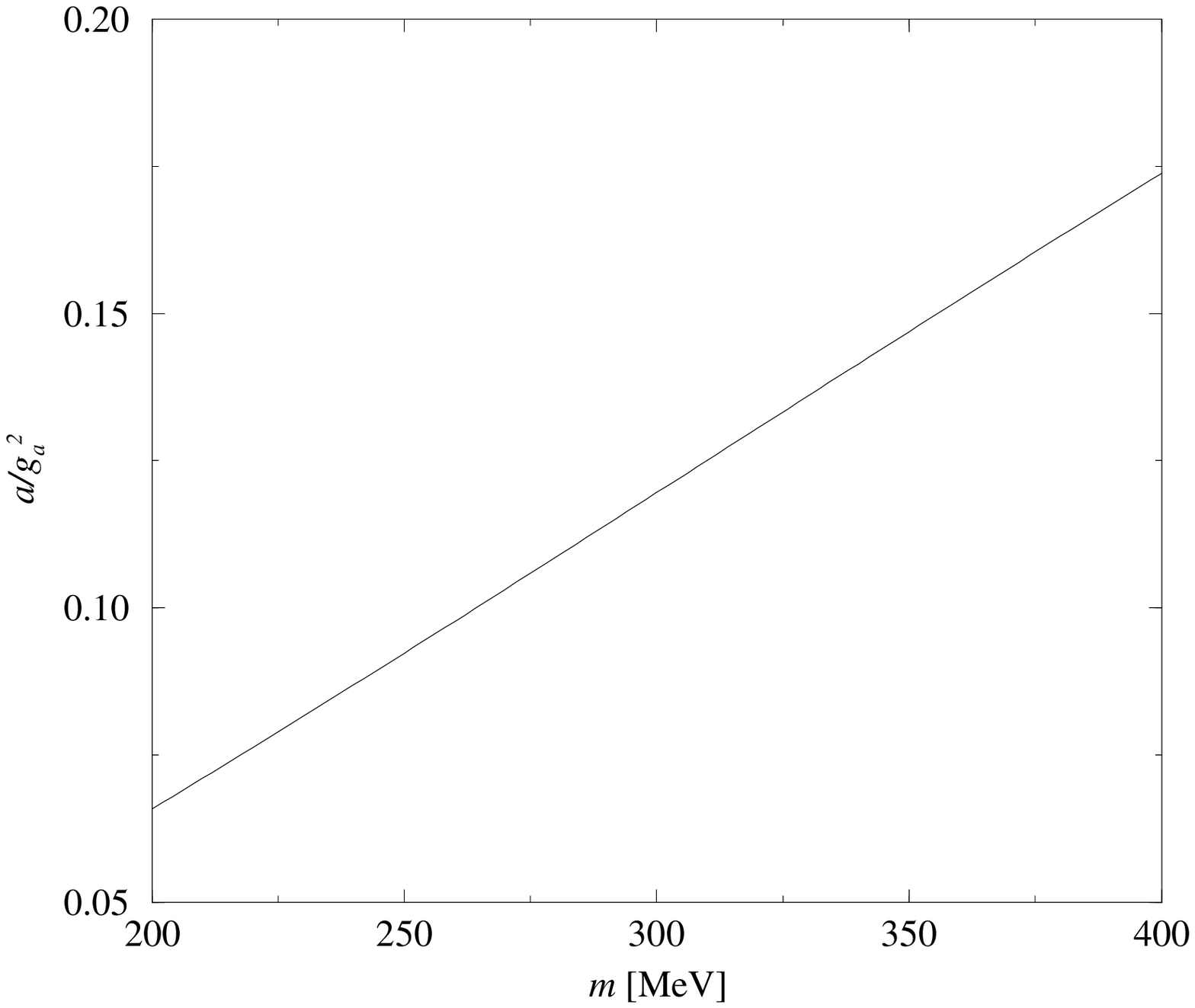,height=7cm}
\caption{The ratio $a/g_a^2$ as a function of the quark mass
$m$. $a/g_a^2 = f(m)$, where $f(m)$ is in integral form. In the
indicated region of $m$, the function $f(m)$ can be fitted with the
linear function $c_0 + c_1 m$, where $c_0 \approx - 0.043$ and $c_1
\approx 0.00054 \, {\rm MeV}^{-1}$}
\label{fig:aga2_200_400}
\end{figure}

It should be noted that the matching parameter $g_a$ and the value of
the parameter $a$ are closely related.  In Fig.~\ref{fig:a_m_ga} we
have displayed $a$ as a function $m$ and $g_a$.  This is further
exemplified in Fig.~\ref{fig:aga2_200_400}, where we show how
$a/g_a^2$ varies with $m$.  We observe that in the relevant region $a$
scales as $g_a^2$ for a fixed value of $m$.  This dependence is
neglected by several authors, who calculate the value of the
axial-vector coupling constant of the nucleon from the difference of
the quark spin polarizations $\Delta q$ as $g_A^{np} = \Delta u -
\Delta d$.  However, when the matching parameter $g_a \neq 1$ the
expression for $g_A^{np}$ must be multiplied by $g_a$, {\it i.e.} $g_A^{np}
= g_a \left(\Delta u - \Delta d\right)$ \cite{mano84,ohls98}.

Consider next the quark spin polarizations. In the $\chi$QM, they are
given by \cite{lind98,chen98}
\begin{eqnarray}
\Delta u &=& \tfrac{4}{3} - a \left(\tfrac{7}{3} + \tfrac{4}{3}\alpha^{2} +
\tfrac{4}{9}\xi' \right),
\label{eq:u}
\\
\Delta d &=& -\tfrac{1}{3} - a \left(\tfrac{2}{3} - \tfrac{1}{3}\alpha^{2} -
\tfrac{1}{9}\xi' \right),
\label{eq:d}
\\
\Delta s &=& - a\alpha^2,
\label{eq:s}
\end{eqnarray}
where $\xi' \equiv \beta^2 + 2 \zeta^2$. Thus for $\beta = 1$ and
$\zeta = 0$, we have $\xi' = 1$.

Using the quark spin polarizations, the observable $g_A^{np}$,
expressed in terms of the $\chi$QM, becomes
\begin{equation}
g_{A}^{np} = g_a \left(\Delta u - \Delta d\right) = \tfrac{5}{3} g_{a}
\left[1 - a \left(1 +
\alpha^{2} + \tfrac{1}{3}\xi' \right)\right].
\label{eq:gapn}
\end{equation}

From our previous discussion, we have seen that the parameters are
compatible with $g_a = 1$, which will be used from now on.  We will
also fix the parameter $\beta$ to $1$, which seems to be favored by
the magnetic moment data \cite{lind98,lind982}, and hence the
parameter $\zeta$ to $0$, which implies that $\xi'=1$.  We have then
only one free parameter $\alpha$ to fit to $g_A^{np} = 1.2601 \pm
0.0025$ \cite{barn96}.  The result is $\alpha \approx 0.54$.  This
can, of course, not be taken as compulsory, since the model would not
allow $g_A^{np}$ to be determined that well.  From magnetic moment
data slightly higher values of $\alpha$, up to $0.7$, are favored
\cite{lind98,lind982}.  Since $\alpha$ is the suppression factor for
kaon GB emission, it can be argued that $\alpha$ is of the order
$m/m_s \simeq 2/3$ \cite{song97,webe97,chen98}.

In Table~\ref{tab:spin_pol} we list the quark spin polarizations
calculated from (\ref{eq:u})--(\ref{eq:s}) with $\xi' = 1$ and
$\alpha \approx 0.54$ and $\alpha = 2/3$, respectively, as well as
some other related quantities.

\section{Nucleon quark spin polarization}
\label{sec:nucleon}

The nucleon quark spin polarizations are usually analyzed by the spin
dependent quark structure functions $g^p_1$ and $g^n_1$.  In these
analyses, the values of $g_{A}^{np}$ and $a_{8}$ are commonly used
\cite{elli93,elli95}.  The experimental value for $a_{8}$ is obtained
from hyperon semileptonic decays using the assumption of SU(3) flavor
symmetry.  However, when the SU(3) symmetry is broken it is not clear
how to connect the axial-vector form factors to the quark spin
polarizations.  Since the parameters in the $\chi$QM can be fixed
without using the hyperon axial-vector form factors, it offers an
independent way to analyze the nucleon spin.  We thus express the
nucleon quark spin polarization directly in terms of the $\chi$QM
quark spin polarizations, avoiding the use of any SU(3) symmetry
assumptions for the value of $a_{8}$.  For our analysis, we will use
the formulas from the analysis of Ellis and Karliner
\cite{elli93,elli95}.

Define the integrals
\begin{eqnarray}
\Gamma^{p}_{1}(Q^2) &\equiv& \int_{0}^{1} g_{1}^{p}(x,Q^{2}) \, dx,
\label{eq:gp}
\\
\Gamma^{n}_{1}(Q^2) &\equiv& \int_{0}^{1} g_{1}^{n}(x,Q^{2}) \, dx,
\label{eq:gn}
\end{eqnarray}
which are the first moments of the proton and neutron spin structure
functions, respectively.  These integrals can be expressed in terms of
the quark spin polarizations, using the evolution equations for
arbitrary $Q^{2}$, by means of two functions, $f = f(Q^{2})$ and $h =
h(Q^{2})$, in the form
\begin{eqnarray}
\Gamma^p(Q^2) &=& \tfrac{1}{9}\Delta u(Q^2) (f+h)
+ \tfrac{1}{18}\Delta d(Q^2) (2h-f) \nonumber \\
&+& \tfrac{1}{18}\Delta s(Q^2) (2h-f),
\label{eq:ggp}
\\
\Gamma^n(Q^2) &=& \tfrac{1}{9}\Delta d(Q^2) (f+h)
+ \tfrac{1}{18}\Delta u(Q^2) (2h-f) \nonumber \\
&+& \tfrac{1}{18}\Delta s(Q^2) (2h-f).
\label{eq:ggn}
\end{eqnarray}
The functions $f$ and $h$ depend on the number of flavors and the
renormalization scheme. In the $\overline{\rm MS}$ scheme for $N_f = 3$
flavors, the functions $f$ and $h$ are given by \cite{lari91}
\begin{eqnarray}
f(\alpha_s(Q^{2})) &=& 1-\frac{\alpha_s(Q^2)}{\pi}-
3.5833\left(\frac{\alpha_s(Q^2)}{\pi}\right)^2 \nonumber \\
&-& 20.2153\left(\frac{\alpha_s(Q^2)}{\pi}\right)^3 \nonumber \\
&-& {\cal O}(130)\left(\frac{\alpha_s(Q^2)}{\pi}\right)^4+\ldots
\end{eqnarray}
and \cite{lari94}
\begin{eqnarray}
h(\alpha_s(Q^{2})) &=& 1-\frac{\alpha_s(Q^2)}{\pi}-
1.0959\left(\frac{\alpha_s(Q^2)}{\pi}\right)^2 \nonumber \\
&-& 3.7\left(\frac{\alpha_s(Q^2)}{\pi}\right)^3+\ldots.
\end{eqnarray}

Ellis and Karliner \cite{elli93,elli95} use the combinations
$g_{A}^{np} \equiv \Delta u - \Delta d = 1.2573 \pm 0.0028$ and $a_{8}
\equiv \Delta u + \Delta d - 2 \Delta s = 0.601 \pm 0.038$ to evaluate
$\Delta\Sigma(Q^{2}) \equiv \Delta u(Q^2) + \Delta d(Q^2) + \Delta
s(Q^2)$. Their most recent value is $\Delta\Sigma(Q^{2}) = 0.31 \pm
0.07$ at a renormalization scale of $Q^2 = 10 \, {\rm GeV^2}$
\cite{elli95}.

However, as mentioned above, rather than using the experimental values
for the axial-vector form factors to evaluate $\Delta\Sigma(Q^{2})$,
we would like to analyze the spin structure integrals directly in
terms of the $\chi$QM.

Expressed in terms of the $\chi$QM parameters, we obtain the following
relations
\begin{eqnarray}
\Gamma^p_1(Q^2) &=& \tfrac{1}{12} f \left(g_A^{np} + \tfrac{1}{3}
a_8\right) + \tfrac{1}{9} h \Delta \Sigma(Q^2) \nonumber \\
&=& \tfrac{1}{6} f \left[1 - a
\left(\tfrac{4}{3} + \tfrac{2}{3}\alpha^{2} + \tfrac{1}{3}\xi'
\right)\right] \nonumber \\
&+& \tfrac{1}{9} h \Delta\Sigma(Q^2),
\label{eq:gggp}
\\
\Gamma^n_1(Q^2) &=& \tfrac{1}{12} f \left(-g_A^{np} + \tfrac{1}{3}
a_8\right) + \tfrac{1}{9} h \Delta \Sigma(Q^2) \nonumber \\
&=& - \tfrac{1}{9} f \left[1 - a \left(\tfrac{1}{2} +
\tfrac{3}{2}\alpha^2 + \tfrac{1}{3}\xi' \right)\right] \nonumber \\
&+& \tfrac{1}{9} h \Delta\Sigma(Q^2).
\label{eq:gggn}
\end{eqnarray}

This is consistent with the fact that in all renormalization schemes
the combinations $g_A^{np} \equiv \Delta u - \Delta d$ and $a_8 \equiv
\Delta u + \Delta d - 2 \Delta s$ are independent of $Q^2$. Above they
appear in the combinations $\pm g_A^{np} + \tfrac{1}{3} a_8$, which
are likewise independent of $Q^2$.

Using the parameter values found in the previous analyses, we can now
extract the value for $\Delta\Sigma(Q^{2})$ from data of
$\Gamma^p_1(Q^2)$ and $\Gamma^n_1(Q^2)$. We use the published world
average values and the Spin Muon Collaboration (SMC) data
\cite{adam97} for this analysis, which is performed at $Q^2 = 5 \,
{\rm GeV}^2$. The results are found in
Table~\ref{tab:DeltaSigmaQ2}. From this table, we can see that the
world average values do not give consistent values for
$\Delta\Sigma(Q^{2})$ from $\Gamma^p_1(Q^2)$ and $\Gamma^n_1(Q^2)$, in
contrast to the SMC measurements.  In what follows, we therefore adopt
the SMC measurements, that give for $\Delta\Sigma(Q^{2})$ the values
$0.292 \pm 0.139$ for $\alpha \approx 0.54$ and $0.287 \pm 0.139$ for
$\alpha = 2/3$. Since these values overlap, we will use the mean value
of them, {\it i.e.}
$$
\Delta\Sigma(Q^{2}) = 0.29 \pm 0.14 \quad \mbox{at $Q^2 = 5 \, {\rm GeV}^2$}.
$$

The total spin of the nucleon in QCD can be decomposed as \cite{jaff90}
\begin{equation}
\frac{1}{2} = \frac{1}{2}\Delta \Sigma +\Delta L_{q} +\Delta g +\Delta
L_{g},
\label{eq:spin}
\end{equation}
where $\Delta\Sigma$ is the spin polarization contribution of the
quarks, $\Delta L_q$ is the orbital angular momentum of the quarks,
$\Delta g$ is the gluon contribution coming from the axial anomaly,
and $\Delta L_g$ is the orbital angular momentum of the gluons.

In the $\chi$QM, the orbital angular momentum of the quarks is
compensated exactly by the spin polarization contribution of the sea
quarks.  Thus, if we write $\Delta\Sigma = \Delta\Sigma_{\rm valence}
+ \Delta\Sigma_{\rm sea}$, we have $\frac{1}{2} \Delta\Sigma_{\rm sea} + \Delta
L_q = 0$ \cite{chen982} and $\Delta\Sigma_{\rm valence}=1$. In the
$\chi$QM, we can therefore interpret the spin of the nucleon as coming
from the valence quarks. The gluonic degrees of freedom are accounted
for essentially by the GBs below $\Lambda_{\chi {\rm SB}}$ and $\Delta
g = \Delta L_g = 0$.

For higher $Q$ values, the $\chi$QM breaks down and we should use
ordinary QCD.  We will therefore assume that the QCD calculation joins
smoothly to the $\chi$QM and interpret the $\Delta\Sigma \equiv \Delta
u + \Delta d + \Delta s$ calculated in the $\chi$QM as the value of
$\Delta\Sigma(Q^{2})$ for $Q^{2}\leq \Lambda_{\chi {\rm SB}}^{2}$. As
$Q^{2}$ increases, this value evolves according to the
Dokshitzer--Gribov--Lipatov--Altarelli--Parisi (DGLAP) evolution
equations \cite{doks77}. In the Adler--Bardeen renormalization scheme
\cite{adle69}, the quark spin polarizations can be written as
\begin{equation}
\Delta q (Q^{2}) = \Delta q - \frac{\alpha_{s}(Q^{2})}{2\pi}\Delta g
(Q^{2}),
\label{eq:deltaqQ2}
\end{equation}
for $q=u,d,s$. This leads to
\begin{eqnarray}
\Delta\Sigma(Q^{2}) &=& \Delta\Sigma
- N_f\frac{\alpha_{s}(Q^{2})}{2\pi}\Delta g(Q^{2}) \nonumber \\
&=& \left[1 - a \left(3 + 2 \alpha^{2} + \tfrac{1}{3}\xi'
\right)\right] \nonumber \\
&-& N_f\frac{\alpha_{s}(Q^{2})}{2\pi}\Delta g(Q^{2}),
\label{eq:delta}
\end{eqnarray}
where $N_f$ is the number of flavors. Here $N_f = 3$. The first term
$\Delta\Sigma$ is interpreted here to be given by the value in the
$\chi$QM. From this it will then be possible to extract $\Delta g$.

With our parameterization, we can calculate $\Delta \Sigma$ from the
$\chi$QM to be $\Delta\Sigma \approx 0.39$ as the mean value for the
two different $\alpha$-values. We can then obtain an estimate of
$\Delta g$ from the value of $\Delta\Sigma(Q^2) = 0.29 \pm 0.14$ at
$Q^2 = 5 \, {\rm GeV}^2$ obtained above. Using $\alpha_{s}(Q^2 = 5 \,
{\rm GeV}^2) = 0.287 \pm 0.020$, which corresponds to $\alpha_s(Q^2 =
M_Z^2) = 0.118 \pm 0.003$ \cite{barn96}, this gives
$$
\Delta g \approx 0.7 \pm 1.0 \quad \mbox{at $Q^2 = 5 \, {\rm GeV}^2$}.
\label{eq:deltag}
$$
This can be compared to the value found by the SMC, which is
$\Delta g = 1.7 \pm 1.1$ at $Q^2 = 5 \, {\rm GeV}^{2}$ \cite{adam97}.

Our analysis of $\Delta g$ does not depend on any assumption of SU(3)
symmetry for the hyperon semileptonic decays.  Nevertheless, the value
$a_8 \approx 0.57$ for $\alpha=2/3$ is in agreement with the value
of $F/D = 0.575 \pm 0.016$\footnote{$F/D = \frac{g_A^{np} + a_8}{3
g_A^{np} - a_8}$} used in the analysis of \cite{abek95}. When the
value of $a_{8}$ changes with $\delta a_{8}$, the change in
$\Delta\Sigma(Q^{2})$ is about $-\tfrac{1}{4}\delta a_{8}$, which
explains the variation in Table~\ref{tab:spin_pol}.

\section{Summary and conclusions}
\label{sec:summary}

In light of the new measurements for the quark sea isospin asymmetry
by the NuSea Collaboration, the broken SU(3) $\chi$QM gives an
excellent fit to the data with the fluctuation parameter $a=0.15$ and
the SU(3) breaking parameters $\alpha \sim 0.6$ and $\beta = 1$, and
the U(3) breaking parameter $\zeta = 0$, corresponding to no $\eta'$
GB.

With these parameter values, the $\chi$QM accounts in a very
satisfactory way for the quark sea isospin asymmetry, the nucleon
quark spin polarizations, the axial-vector form factor of the nucleon,
the magnetic moments of the nucleons and hyperons plus many other
features of low energy hadron physics, such as the nucleon deep
inelastic scattering structure functions.  In addition, the new data
are fully consistent with the matching parameter $g_{a}=1$, as
suggested by the analysis of Weinberg \cite{wein90,wein91}.  (However,
see also \cite{peri93} for an extended discussion of possible
values for $g_{a}$.)

The $\chi$QM also offers an independent way of analyzing the nucleon
spin problem, relying only on nucleon data. This leads eventually to
an estimate of $\Delta \Sigma(Q^2) = 0.29 \pm 0.14$ at $Q^2 = 5 \,
{\rm GeV}^2$ and $\Delta g \approx 0.7 \pm 1.0$ for the gluon spin
polarization in the nucleon.

\begin{acknowledgement}
{\it Acknowledgements.} This work was supported by the Swedish Natural
Science Research Council (NFR), Contract No. F-AA/FU03281-311. Support
for this work was also provided by the Engineer Ernst Johnson
Foundation (T.O.).
\end{acknowledgement}

\twocolumn

\begin{table}
\caption{The quark spin polarizations and $a_i$, where $i=0,3,8$. $a_0
= \Delta \Sigma \equiv \Delta u + \Delta d + \Delta s$, $a_3 =
g_A^{np} \equiv \Delta u - \Delta d$, and $a_8 \equiv \Delta u +
\Delta d - 2 \Delta s$. The experimental values from
\protect\cite{elli95} have been obtained assuming $g_A^{np} =
1.2573 \pm 0.0028$ and $a_8 = 0.601 \pm 0.038$ and the experimental
values from \protect\cite{abek95} have been obtained assuming
$g_A^{np} = 1.2573 \pm 0.0028$ and $F/D = 0.575 \pm 0.016$. The data
for the $\chi$QM are calculated using $a = 0.15$}
\begin{tabular}{lcccc}
\hline
Quantity & Experimental & NQM & $\chi$QM & $\chi$QM\\
 & value & & $\alpha \approx 0.54$ & $\alpha = \frac{2}{3}$\\
\hline
$\Delta u$ & $0.83 \pm 0.03$ \protect\cite{elli95} & $\frac{4}{3}$ &
 $0.86$ & $0.83$\\
$\Delta d$ & $-0.43 \pm 0.03$ \protect\cite{elli95} & $-\frac{1}{3}$ &
 $-0.40$ & $-0.39$\\
$\Delta s$ & $-0.10 \pm 0.03$ \protect\cite{elli95} & $0$ & $-0.04$ & $-0.07$\\
 & $-0.09 \pm 0.02$ \protect\cite{abek95}\\
$a_0$ & $0.31 \pm 0.07$ \protect\cite{elli95} & $1$ & $0.41$ & $0.37$\\
 & $0.30 \pm 0.06$ \protect\cite{abek95}\\
$a_3$ & $1.2601 \pm 0.0025$ \protect\cite{barn96} & $\frac{5}{3}$ &
$1.2601$ (input) & $1.22$\\
$a_8$ & $0.601 \pm 0.038$ \protect\cite{hsue88} & $1$ & $0.54$ & $0.57$\\
\hline
\end{tabular}
\label{tab:spin_pol}
\end{table}

\begin{table}
\caption{The total quark spin polarization $\Delta\Sigma(Q^2)$ at $Q^2
= 5 \, {\rm GeV}^2$. The data for the $\chi$QM are calculated using
$\alpha_s(Q^2 = 5 \, {\rm GeV}^2) = 0.287 \pm 0.020$ (corresponding to
$\alpha_s(Q^2 = M_Z^2) = 0.118 \pm 0.003$ \cite{barn96})}
\begin{tabular}{ccccc}
\hline
 & \multicolumn{2}{c}{Experiments} & \multicolumn{2}{c}{$\Delta\Sigma(Q^2)$}\\
\hline
\multicolumn{2}{c}{} & Experimental & $\chi$QM & $\chi$QM\\
\multicolumn{2}{c}{} & value & $\alpha \approx 0.54$ & $\alpha = \frac{2}{3}$\\
\hline
 & SMC & $0.132 \pm 0.017$ & $0.295 \pm 0.171$ & $0.317 \pm 0.171$\\
$\Gamma^p_1$\\
 & World average & $0.142 \pm 0.011$ & $0.395 \pm 0.110$ & $0.417 \pm
0.110$\\
\hline
 & SMC & $-0.048 \pm 0.022$ & $0.289 \pm 0.221$ & $0.257 \pm 0.221$\\
$\Gamma^n_1$\\
 & World average & $-0.061 \pm 0.016$ & $0.156 \pm 0.161$ & $0.126 \pm
0.161$\\
\hline
\end{tabular}
\label{tab:DeltaSigmaQ2}
\end{table}

\twocolumn

\begin{thebibliography}{}
\bibitem{mano84} A. Manohar and H. Georgi, Nucl. Phys. \textbf{B234}, 189
(1984)

\bibitem{eich92} E.J. Eichten, I. Hinchliffe, and C. Quigg,
Phys. Rev. D \textbf{45}, 2269 (1992)

\bibitem{chen95} T.P. Cheng and L.-F. Li,
Phys. Rev. Lett. \textbf{74}, 2872 (1995)

\bibitem{song97} X. Song, J.S. McCarthy, and H.J. Weber, Phys. Rev. D
\textbf{55}, 2624 (1997)

\bibitem{webe97} H.J. Weber, X. Song, and M. Kirchbach,
Mod. Phys. Lett. A \textbf{12}, 729 (1997)

\bibitem{webe972} H.J. Weber and K. Bodoor, Int. J. Mod. Phys. E \textbf{6},
693 (1997)

\bibitem{chen98} T.P. Cheng and L.-F. Li, Phys. Rev. D \textbf{57}, 344 (1998)

\bibitem{lind98} J. Linde, T. Ohlsson, and H. Snellman, Phys. Rev. D
\textbf{57}, 452 (1998)

\bibitem{song98} X. Song, Phys. Rev. D \textbf{57}, 4114 (1998)

\bibitem{lind982} J. Linde, T. Ohlsson, and H. Snellman, Phys. Rev. D
\textbf{57}, 5916 (1998)

\bibitem{chen982} T.P. Cheng and L.-F. Li,
Phys. Rev. Lett. \textbf{80}, 2789 (1998)

\bibitem{webe98} H.J. Weber, Mod. Phys. Lett. A \textbf{13}, 71 (1998)

\bibitem{ohls98} T. Ohlsson and H. Snellman, hep-ph/9803490,
Eur. Phys. J. C (to be printed)

\bibitem{song982} X. Song, INPP-UVA-97-08, hep-ph/9802206

\bibitem{song983} X. Song, INPP-UVA-98-03, hep-ph/9804461

\bibitem{bald94} NA51 Collaboration, A. Baldit {\it et al.}, Phys.
Lett. B \textbf{332}, 244 (1994)

\bibitem{amau91} New Muon Collaboration, P. Amaudruz {\it et al.},
Phys. Rev. Lett. \textbf{66}, 2712 (1991)

\bibitem{arne94} New Muon Collaboration, M. Arneodo {\it et al.},
Phys. Rev. D \textbf{50}, R1 (1994)

\bibitem{hawk98} NuSea Collaboration, E.A. Hawker {\it et al.},
Phys. Rev. Lett. \textbf{80}, 3715 (1998)

\bibitem{peng98} NuSea Collaboration, J.C. Peng {\it et al.},
Phys. Rev. D \textbf{58}, 092004 (1998)

\bibitem{wein90} S. Weinberg, Phys. Rev. Lett. \textbf{65}, 1181 (1990)

\bibitem{wein91} S. Weinberg, Phys. Rev. Lett. \textbf{67}, 3473 (1991)

\bibitem{thoo76} G. 't Hooft, Phys. Rev. Lett. \textbf{37}, 8 (1976);
Phys. Rev. D \textbf{14}, 3432 (1976)

\bibitem{gott67} K. Gottfried, Phys. Rev. Lett. \textbf{18}, 1174 (1967)

\bibitem{barn96} Particle Data Group, R.M. Barnett {\it et al.}, {\it
Review of Particle Physics}, Phys. Rev. D \textbf{54}, 1 (1996)

\bibitem{elli93} J. Ellis and M. Karliner, Phys. Lett. B \textbf{313},
131 (1993)

\bibitem{elli95} J. Ellis and M. Karliner, Phys. Lett. B \textbf{341},
397 (1995)

\bibitem{lari91} S.A. Larin, F.V. Tkachev, and J.A.M. Vermaseren,
Phys. Rev. Lett. \textbf{66}, 862 (1991); S.A. Larin and
J.A.M. Vermaseren, Phys. Lett. B \textbf{259}, 345 (1991); A.L. Kataev
and V. Starshenko, Mod. Phys. Lett. A \textbf{10}, 235 (1995)

\bibitem{lari94} S.A. Larin, Phys. Lett. B \textbf{334}, 192 (1994);
A.L. Kataev, Phys. Rev. D \textbf{50}, R5469 (1994)

\bibitem{adam97} Spin Muon Collaboration, D. Adams {\it et al.},
Phys. Rev. D \textbf{56}, 5330 (1997)

\bibitem{jaff90} R.L. Jaffe and A. Manohar, Nucl. Phys. \textbf{B337},
509 (1990)

\bibitem{doks77} Y.L. Dokshitzer, Sov. Phys. JETP \textbf{46}, 461 (1977);
V.N. Gribov and L.N. Lipatov, Sov. J. Nucl. Phys. \textbf{15}, 438 (1972);
675 (1972); G. Altarelli and G. Parisi, Nucl. Phys. \textbf{B126}, 298 (1977)

\bibitem{adle69} S. Adler and W. Bardeen, Phys. Rev. \textbf{182}, 1517 (1969);
G. Altarelli and G.C. Ross, Phys. Lett. B \textbf{212}, 391 (1988)

\bibitem{abek95} E143 Collaboration, K. Abe {\it et al.},
Phys. Rev. Lett. \textbf{75}, 25 (1995)

\bibitem{peri93} S. Peris and E. de Rafael, Phys. Lett. B
\textbf{309}, 389 (1993)

\bibitem{hsue88} S.Y. Hsueh {\it et al.}, Phys. Rev. D \textbf{38}, 2056 (1988)

\end{thebibliography}
\end{document}